\documentstyle[preprint,pre,eqsecnum,aps]{revtex}
\begin{document}
\draft
\title{%
Orbits in one-dimensional finite linear cellular automata}
\author{Shin-ichi Tadaki\cite{email}}
\address{Department of Information Science, 
Saga University, Saga 840, Japan}
\date{\today}
\maketitle
\begin{abstract}%
Periodicity and relaxation are investigated for the
trajectories of the states in one-dimensional finite cellular automata
with rule-90 and 150.
The time evolutions are described with matrices.
Eigenvalue analysis is applied to clarify the maximum value of
period and relaxation.
\end{abstract}
\narrowtext
\section{Introduction}
Cellular Automata are one of the simplest mathematical models for 
nonlinear dynamics to produce complicated patterns of behaviour,
which had been  originally introduced by von Neumann\cite{Neumann}.  
Wolfram had reintroduced cellular automata as a
model to investigate complexity and randomness\cite{Wolfram:1}.  
He investigated many fundamental 
features of them\cite{Wolfram:2,Wolfram:3,Martin}.
Since then many authors have made efforts to clarify the properties 
of cellular automata and applied to natural systems\cite{Gutowitz}.

One-dimensional cellular automata are described by the
discrete time evolution of site $a_i$:
\begin{equation}
a_i(t+1) = F[a_{i-r}(t), a_{i-r+1}(t),\ldots,a_i(t),\ldots,a_{i+r}(t)],
\end{equation}
where $a_i$ takes $k$ discrete values over $Z_k$.
The simplest model, {\em elementary cellular automaton}, consists of sites 
with two internal states over $Z_2$ interacting with the nearest
neighbour sites ($r=1$).
Wolfram introduced a naming scheme for these models and
classified the behaviour of cellular automata into four 
classes\cite{Wolfram:1,Wolfram:2}.

Most authors have worked on cellular automata within the scope of the
infinite number of sites.
A few works have concerned periodic boundary condition
(cylindrical automata)\cite{Martin,Wolfram:4,Jen,Voorhees,Guan}.
In our previous paper\cite{Tadaki} (referred as paper I)
we had investigated the periodic orbits of finite rule-90 cellular 
automata with Dirichlet boundary condition.
We analyzed the eigenvalue equations of the transfer matrices
which describe the time evolution of the system.
In the present paper the method is extended to the rule-150 case.
Some results obtained in paper I will be cited again for completeness.

\section{Models and numerical results}
We review the numerical results obtained in paper I on so-called 
rule-90 cellular automata following Wolfram's naming scheme.
The time evolution of the $i$-the site $a_i(t)\in\{0,1\}$ $(i=1\sim N)$
is described as a sum modulo 2 of the nearest neighbour sites:
\begin{equation}
a_i(t+1)=a_{i-1}(t)+a_{i+1}(t)\bmod2.
\end{equation}
We use the Dirichlet boundary conditions, $a_0=a_{N+1}=0$.
In the Wolfram's classification the rule-90 model
belongs to the third class which shows the chaotic behaviour.
The time evolution is also expressed by matrix:
\begin{equation}
A(t+1)=UA(t),
\end{equation}
where $A(t) =\ ^t(a_1(t),a_2(t),\ldots,a_N(t))$ describes the
state at $t$ and the transfer matrix $U$ is given by
\begin{equation}
U_{ij}=\cases{1&$j=i\pm1$,\cr 0&otherwise.\cr}
\end{equation}
In paper I we had found the periodic structures of the transfer 
matrices $U$ numerically as shown in Table~\ref{periodtable}.
They are summarized as follows:
\begin{mathletters}
\begin{eqnarray}
&U^{\Pi_N}=I,&\quad(\hbox{$N$ is even}),\label{even90}\\
&U^{\Pi_N+\pi_N}=U^{\pi_N},&\quad(\hbox{$N$ is odd}),\label{odd90}\\
&U^N=0,&\quad(N=2^n-1).\label{special90}
\end{eqnarray}
\end{mathletters}%

For cellular automata with even number of sites (Eq.~(\ref{even90})),
every state except the null one (all sites are zero) is
on the orbits with period not exceeding $\Pi_N$.
The concrete periods depend on the initial states.
The period $\Pi_N$ corresponds to the least common multiple of them,
and we call it the maximum period.
The states on the orbits with the maximum period have the lowest symmetry.
The states with some symmetries are on the orbits with shorter periods
than $\Pi_N$.
For cellular automata with odd number of sites except $N=2^n-1$
(Eq.~(\ref{odd90})),
some states belong to the orbits with period $\Pi_N$ or less
and the others except the null state are drawn to the periodic orbits
after some time steps not exceeding the maximum relaxation $\pi_N$.
It is very interesting that in the $N=2^n-1\ (n\in Z)$ case
every state is drawn into the null state after at most $N$ steps
(Eq.~(\ref{special90})).
In this case the configuration space has only one
basin with the null state at the center.
Schematic features of trajectories are shown in Fig~\ref{bigschematic}.

Next we investigate the rule-150 cellular automata,
which also belongs to the third class in the Wolfram's classification.
The time evolution is described as a sum modulo 2 of the
site itself and the nearest neighbors:
\begin{equation}
a_i(t+1)=a_{i-1}(t)+a_i(t)+a_{i+1}(t)\bmod2.
\end{equation}
The transfer matrix is given by
\begin{equation}
U_{ij}=\cases{1&$j=i\pm1$ or $j=i$,\cr
              0&otherwise.\cr}
\end{equation}
The periodicity of the transfer matrices are also found numerically
as shown in Table~\ref{periodtable}.
There are no cases drawn into the null state as described by
Eq.~(\ref{special90}).
For cellular automata with $N=3n+2$ there appear the
periodic orbits of period not exceeding $\Pi_N$ with the relaxation path 
whose maximum length is $\pi_N$.
For the others $N\not=3n+2$ all states are on the periodic orbits
except the null state.

The maximum periods $\Pi_N$ of the rule-90 and rule-150 cellular automata 
are found to coincide each other in many cases of the number of sites
(See Table~\ref{periodtable}, Figs.~\ref{Period90} and \ref{Period150}).
Grassberger had reported the similarity
between the rule-90 and rule-150 cases in behaviour\cite{Grassberger}.
Especially the steepest peaks $N=6,10,18,22,28$
coincide between the rule-90 and rule-150 cases,
whose periods are expressed by $\Pi_N=2^{N/2+1}-2$.
The maximum periods grow exponentially
for the average of all number of sites except $N=2^n-1$ cases,
in which all state are drawn into the null states within $N$ steps for
rule-90 and $\Pi_N=N+1$ for rule-150.

\section{Eigenvalue analysis}
The trajectories of rule-90 cylindrical automata 
had been investigated by Martin, Odlyzko and Wolfram\cite{Martin}.
They analyzed {\it characteristic polynomials\/}
which describe the states of cellular automata.
The method seems to work well in the periodic boundary cases.
To investigate the periodic structures analytically in the Dirichlet
boundary cases,
we had introduced the simpler method to analyze the eigenvalue
polynomials in paper I.
The eigenvalue equation of the transfer matrix
\begin{equation}
UA = - \lambda A,
\end{equation}
reads the secular equation:
\begin{equation}
D^N(\lambda)\equiv\vert U+\lambda I\vert = 0.
\end{equation}
All calculation are carried out
over $Z_2$ since a site $a_i$ takes binary values.

The recursion relation of the eigenvalue polynomials for rule-90
cellular automata
\begin{equation}
D^N(\lambda) = \lambda D^{N-1}(\lambda) - D^{N-2}(\lambda),\label{rec90}
\end{equation}
gives the explicit form of $D^N(\lambda)$.
The eigenvalue polynomial $D^N(\lambda)$ is over Galois field 
of order 2, $GF(2)$, namely all coefficients of $\lambda^i$
are over $Z_2$:
\begin{equation}
D^N(\lambda) = 
\sum_{j=0}^{j_{\rm max}}\left(C^N_j\bmod2\right)\lambda^{N-2j},
\label{polynomial}
\end{equation}
where $j_{\rm max}=\lfloor N/2\rfloor$, $\lfloor\ \rfloor$ is a Gaussian
symbol and
\begin{equation}
C^N_j\equiv (-1)^j{N-j\choose j}.
\end{equation}
Note that the definition of $C^N_j$ is slightly different from that in
paper I.

The eigenvalue polynomials are given by the replacement 
$\lambda\rightarrow\lambda+1$ in Eqs.~(\ref{rec90}) and 
(\ref{polynomial}) for the rule-150 cellular automata:
\begin{equation}
D^N(\lambda) = (\lambda+1) D^{N-1}(\lambda) - D^{N-2}(\lambda),
\label{rec150}
\end{equation}
\begin{eqnarray}
D^N(\lambda) 
&=& \sum_{j=0}^{j_{\rm max}}\left(C^N_j\bmod2\right)(\lambda+1)^{N-2j},
\nonumber\\
&=& \sum_{k=0}^N\sum_{j=0}^{\lfloor(N-k)/2\rfloor}
\left(C'^N_{j,k}\bmod2\right)\lambda^k,\\
&&C'^N_{j,k}\equiv(-1)^j{N-j\choose j}{N-2j\choose k}.
\end{eqnarray}
Some examples of $D^N(\lambda)$ for the rule-90 and rule-150 cellular
automata are shown in Tables~\ref{polynomial90} and \ref{polynomial150}.

These eigenvalue equations enable us to find
the maximum periods and maximum length of relaxation path to the 
periodic orbits.
First we study nilpotent cases, $D^N(\lambda)=\lambda^{P(N)}$.
In paper I we found that those happen for the rule-90 with 
$N=2^n-1$ sites and $P(N)=N$.
All states are drawn into the null state within steps $N$ or less.
The rule-150 model does not show the similar behaviour.

Next we discuss the cases that there are constant terms in
polynomials.  Those happen for the rule-90 with even number of sites
since $C^N_{j_{\rm max}}=(-1)^{N/2-2}$ for even $N$.
For the rule-150 case the polynomials has a constant term
for $N\not=3n+2$ by Eq.~(\ref{rec150}).
The eigenvalue polynomials reduce to the simple form as 
$\lambda^{P(N)}+1=0$ by multiplying some power of
$\lambda$ and repeatedly substituting the eigenvalue equation to
itself\cite{num,numnum}.  The minimum value of the
power $P(N)$ corresponds to the maximum period $\Pi_N$.  If
the eigenvalue polynomial is factorized, we are able to get
shorter periods depending on the initial state from those
factors by the procedure mentioned above.  For instance, we
consider $N=4$ rule-150 cellular automata (Fig.~\ref{orbits4}).  
The eigenvalue equation $\lambda^4+\lambda^2+1=0$ reduces to $\lambda^6+1=0$
and the maximum period is $\Pi_4=6$.  
The eigenvalue equation is also factorized to $(\lambda^2+\lambda+1)^2=0$.
We find another solution $\lambda^3=1$ from $\lambda^2+\lambda+1=0$.
Therefore there are period 6 and period 3 orbits for $N=4$ rule-150 
cellular automata.

Explicit expression of the maximum period $\Pi_N$ is simply obtained
for $N=2^n-1$ rule-150 cellular automata.
In this case the identity ${N-j\choose j}\bmod2=0$ holds for $j\not=0$.
The non-zero elements of $C'^N_{j,k}$ are $C'^N_{0,k}={N\choose k}$.
By the identity ${2^n-1\choose k}\bmod2=1$ for all non-negative
integers $k$, the eigenvalue polynomial reduces to
$D^N(\lambda)=\sum_{k=0}^{N}\lambda^k$.
Following the above procedure, we find the maximum period
$\Pi_N=N+1$.

Finally we study the cases that the lowest powers of polynomials
are greater than 0, namely they have the forms as 
$D^N(\lambda)=\lambda^{p(N)}\cdot f(\lambda)$,
where $f(\lambda)$ is a polynomial with a constant term.
The number of null solutions, $p(N)$, corresponds to the
maximum length of relaxation path $\pi_N$.
Applying the above procedure to $f(\lambda)$,
we obtain the maximum period and shorter ones.
For example, we show the case of $N=5$ rule-90 cellular automata 
(Fig.~\ref{orbits5}). The eigenvalue polynomials is
$D^N(\lambda) = \lambda\cdot(\lambda^4+1)$.
The maximum length of relaxation path is $\pi_5=1$. From $\lambda^4+1=0$, 
we find the maximum period $\Pi_5=4$.
We also find shorter ones 2 and 1 by factorizing
$\lambda^4+1=(\lambda^2+1)^2=(\lambda+1)^4$.

A simple recursion relation of the maximum period and maximum
length of relaxation are found for rule-90 cellular automata 
with odd number of sites.
The new recursion relation
\begin{equation}
D^N(\lambda) = \lambda^2D^{N-2}(\lambda) - D^{N-4}(\lambda),
\end{equation}
holds by virtue of algebra in $GF(2)$.
This gives a simple form  $D^{2n+1}(\lambda)=\lambda D^n(\lambda^2)$.
The maximum period and the maximum length of relaxation path of 
$N=2n+1$ rule-90 cellular automata are described by those of
$N=n$ sites as
\begin{eqnarray}
\Pi_{2n+1}&=&2\Pi_n+1,\\
\pi_{2n+1}&=&2\pi_n+1.
\end{eqnarray}

\section{Concluding remarks}
We investigate the periodic structure of one-dimensional rule-90
and rule-150 cellular automata with Dirichlet boundary condition.
We find three types of behaviour.
The first is periodic one which appears in cellular automata
with even number of sites for rule-90 and
$N\not=3n+2$ for rule-150.
The second appears in the cases of odd number of sites for rule-90
and $N=3n+1$ for rule-150.
There are some periodic orbits and irreversible relaxation paths to them.
The peculiar behaviour happens to the case $N=2^n-1$ for rule-90.
All states are drawn into the null state within $N$ steps.

The eigenvalue equations of the transfer matrices are analyzed.
The maximum period is obtained by finding the minimum power to
satisfy $\lambda^{P(N)}+1=0$.
Shorter periods are given by factorizing the eigenvalue polynomials.
The number of null solutions of the polynomials gives the
maximum length of relaxation path.
For some special cases, we find the explicit forms of the maximum period.
Distribution of periods and relaxation is still not clear.

Roots of the eigenvalue equations, in general, are found
over the Galois extension of finite field\cite{num,numnum}.
For all positive integer $N$ there are primitive
polynomials ${\cal P}(x)$ of degree $N$ over $GF(2)$.
One of the roots, $\alpha$, of ${\cal P}(x)$ generates the 
Galois extension $GF(2^N)$, whose elements are 
$\lbrace 0,1,\alpha,\alpha^2,\ldots,\alpha^{2^N-2}\rbrace$
and $\alpha^{2^N-1}=1$.
Other roots of ${\cal P}(x)$ are called {\it conjugate\/} of $\alpha$
and generate the {\it isomorphic\/} Galois extensions.
A general polynomial of degree $N$ has roots over $GF(2^N)$.
A eigenvalue polynomial has $N$ roots 
$\lbrace a_i\rbrace\in GF(2^N)\ (i=1,2,\ldots,N)$.
For instance $N=4$ rule-150 case, the equation $\lambda^4+\lambda^2+1=0$
has roots in $\lbrace 0,1,\alpha,\alpha^2,\ldots,\alpha^{14}\rbrace\ 
(\alpha^{15}=1)$, where $\alpha$ is one of roots of forth
primitive polynomial $\lambda^4+\lambda+1=0$.
Explicitly the roots are $\alpha^5$ and $\alpha^{10}$ with
multiplicity 2 for each.
The minimum common multiplier of the roots, which satisfies $\alpha^k=1$,
is $\alpha^{30}$.
This seems to suggest the maximum period $6$,
namely $(\alpha^5)^6=(\alpha^{10})^3=\alpha^{30}=1$.
This procedure, however, does not work for other cases.
For example, the eigenvalue polynomial of $N=3$ rule-150 case
is $\lambda^3 + \lambda^2+\lambda+1$.
It can be factorized to $(\lambda+1)^3$ and the root is $\lambda=1$ with
multiplicity 3.
The periods, however, are 4, 2 and 1.
More number theoretical studies are expected.

The elementary cellular automata are also a subject to
build a built-in self-test of VLSI\cite{Hortensius,Compagner}.
Usually the shift-registers which generate
pseudo-random sequences of length $2^N-1$ with $N$ registers,
are used for build-in self-tests.
Our results show the elementary cellular automata can also generate
exponentially long but not maximum sequences.
By fine-tuned mixture of rule-90 and rule-150 cases, 
{\it hybrid cellular automata}, it is shown to be able to produce
a maximum length pseudo-random sequences.

\acknowledgments
The author would like to thank S.~Matsufuji for
valuable discussions and comments.


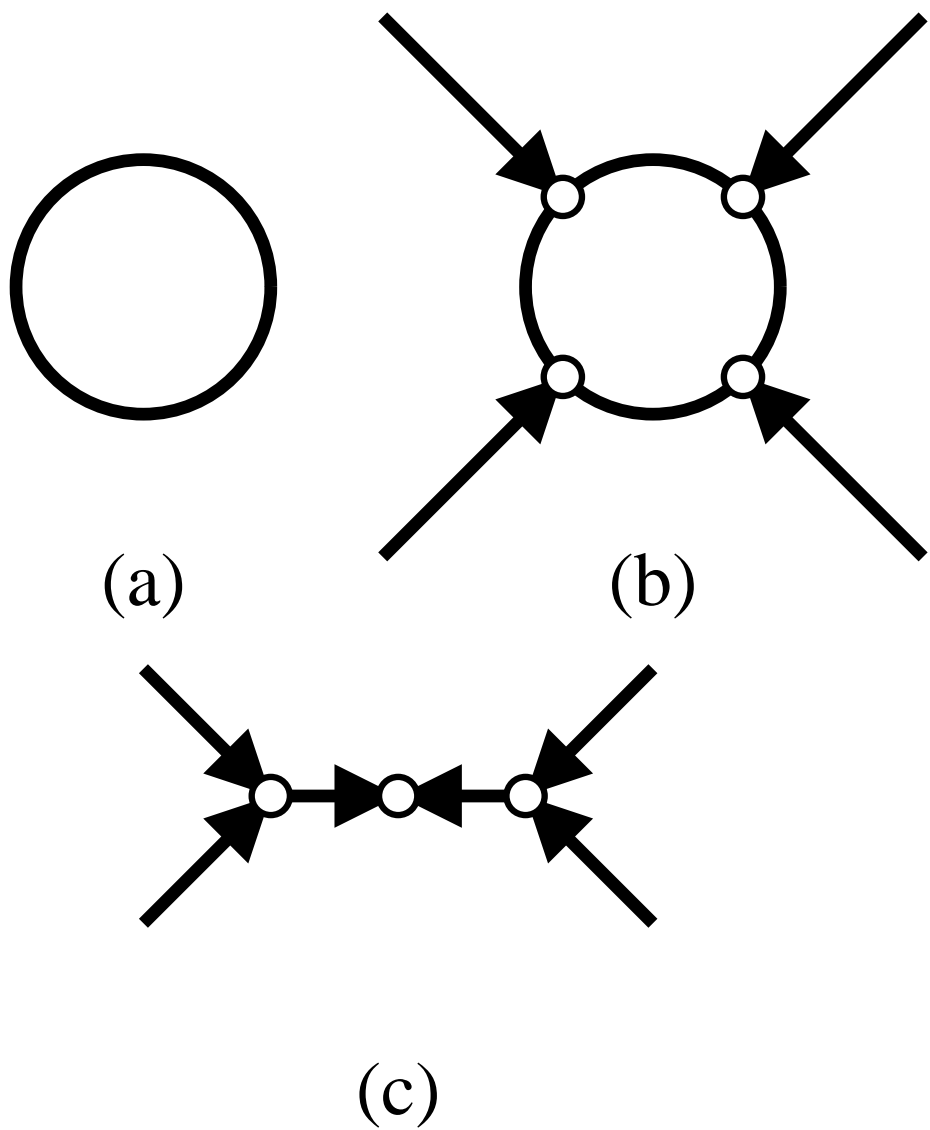
\begin{figure}
\caption{Schematic features of the trajectories of cellular automata:
(a) simple periodic orbit, (b) periodic orbit with relaxation,
(c) limit point.}
\label{bigschematic}
\end{figure}
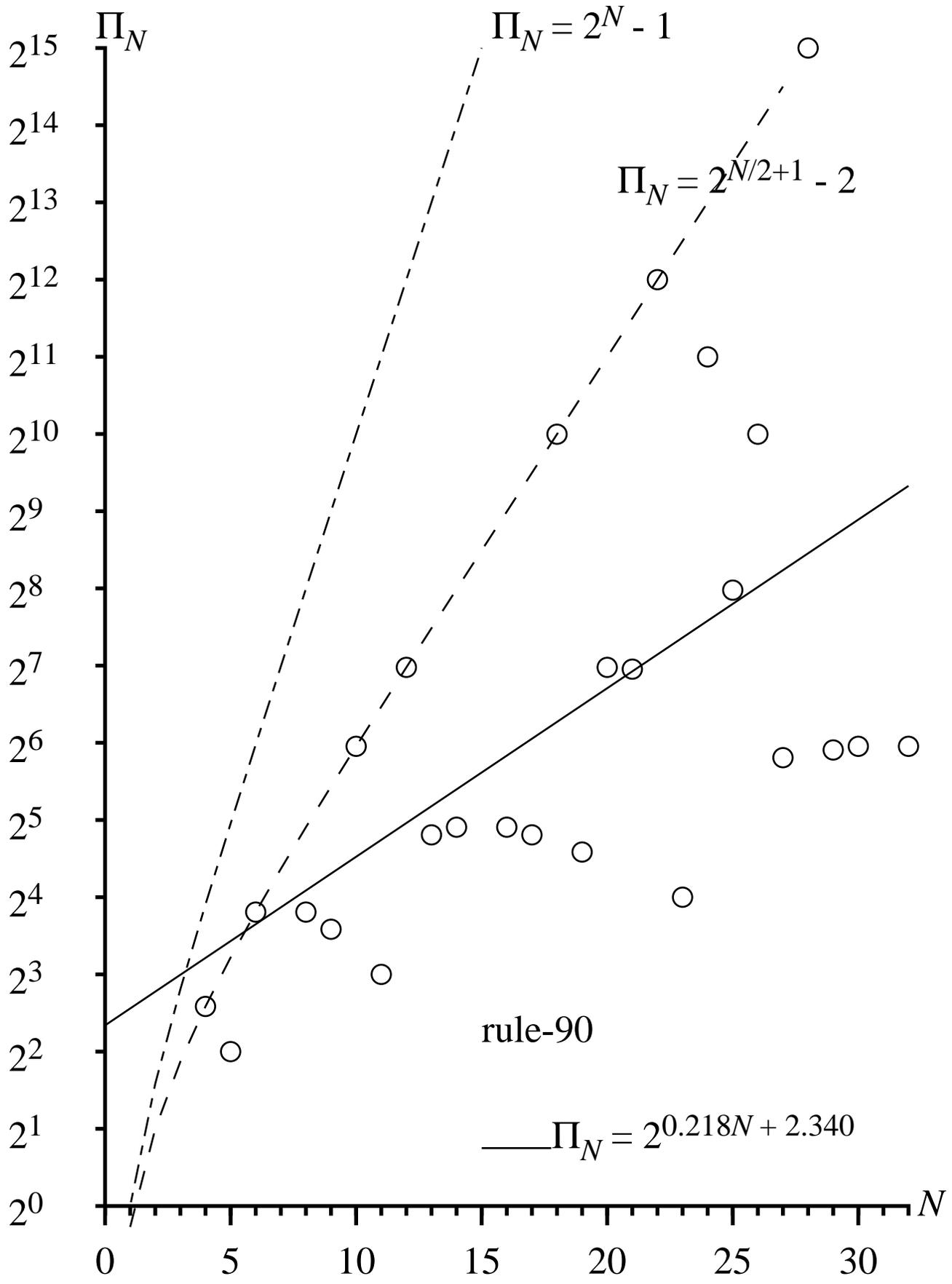
\begin{figure}
\caption{%
Distribution of the maximum periods of the rule-90 cellular automata.
The bold line is the average of the periods except $N=2^n-1$ cases.
The broken one denotes the curve $\Pi_N=2^{N/2+1}-2$,
which fits peaks of $N=6,10,18,22,28$.}
\label{Period90}
\end{figure}
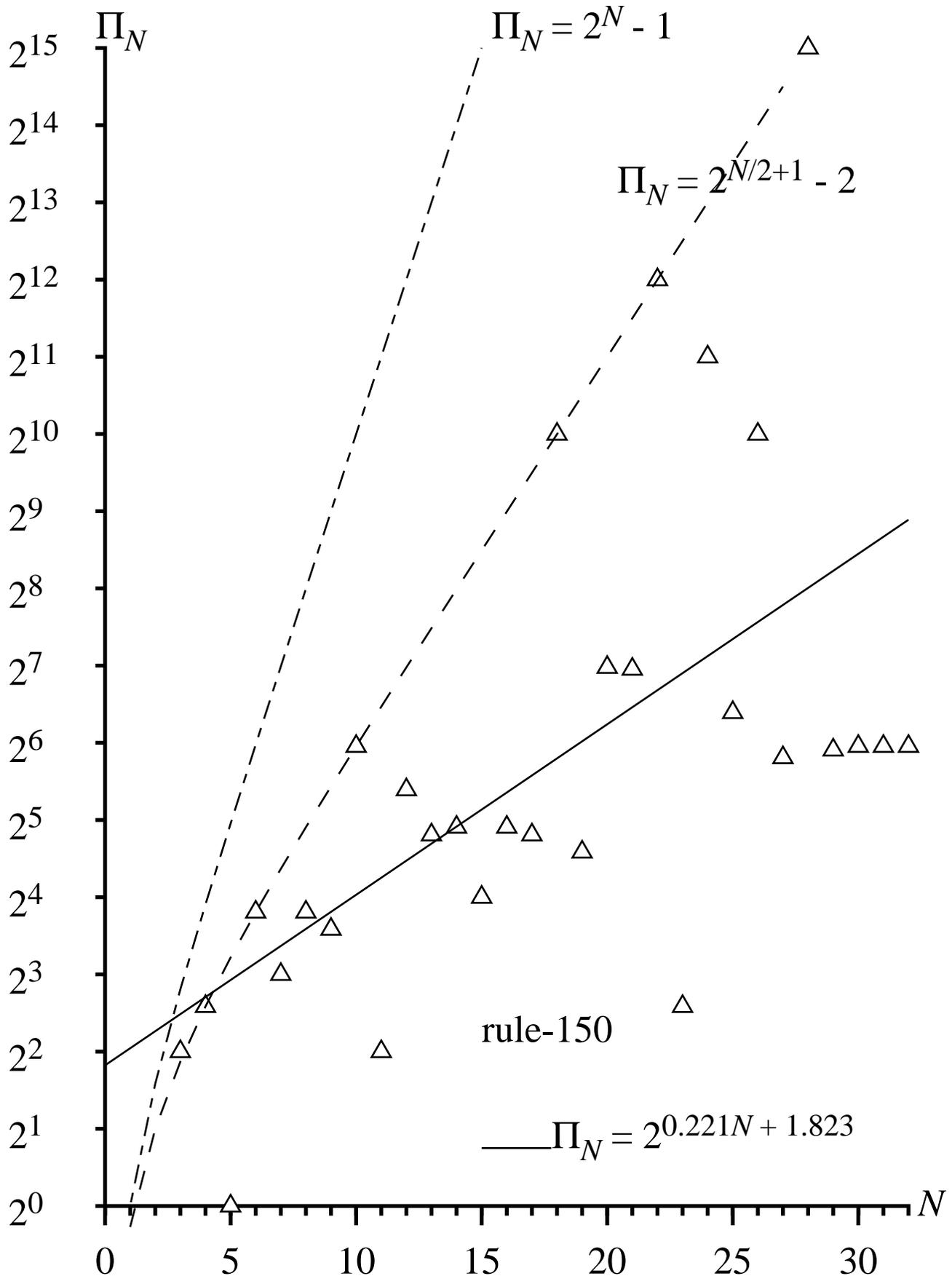
\begin{figure}
\caption{%
Distribution of the maximum periods of the rule-150 cellular automata.
The bold line is the average of the periods of all number of sites.
The broken one denotes the same as in Fig.~\protect{\ref{Period90}}.
}
\label{Period150}
\end{figure}
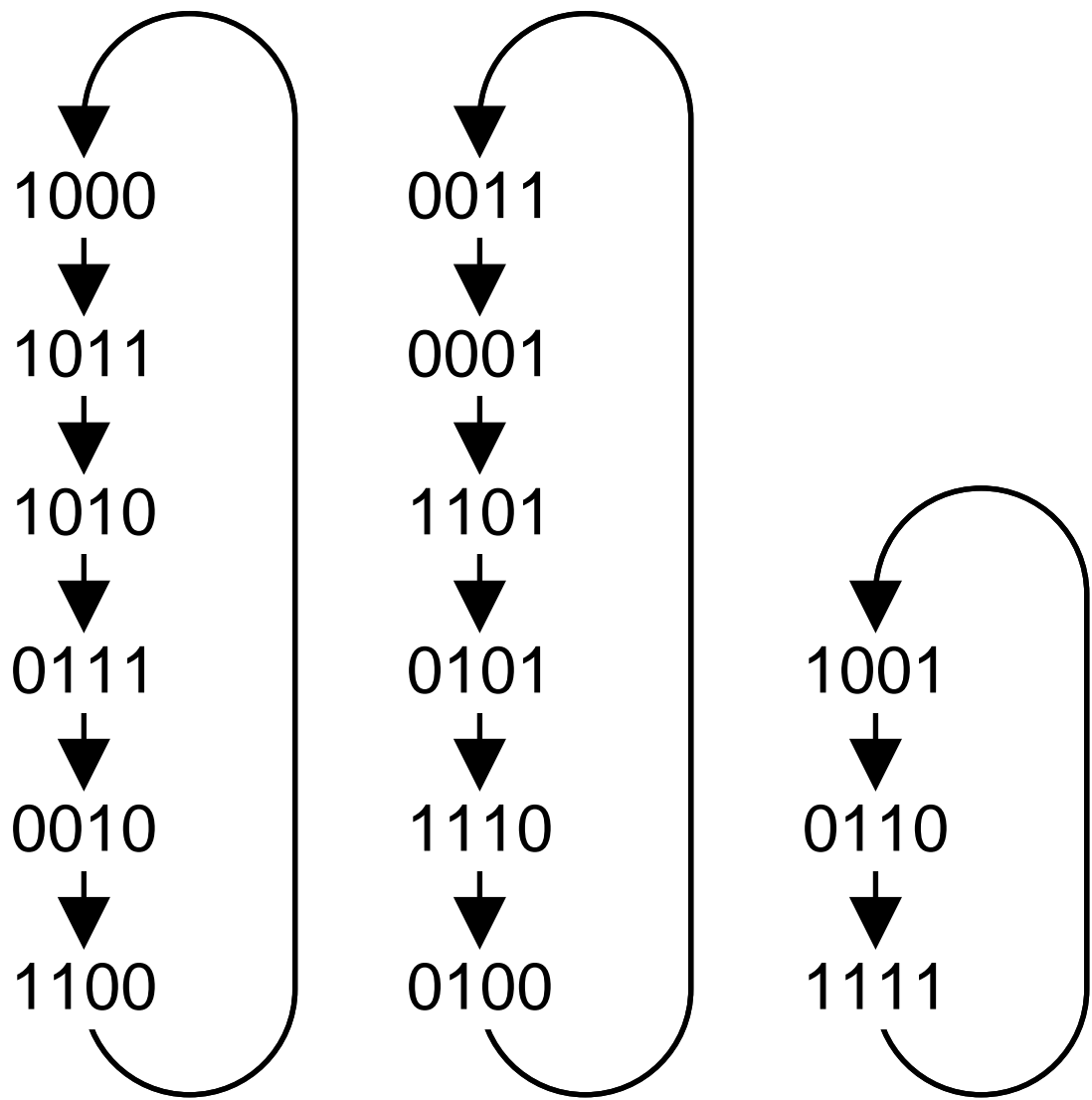
\begin{figure}
\caption{The orbits of $N=4$ rule-150 cellular automaton.
All states are classified into 3 orbits excepts the null states.
Two of them are period 6 and the other period 3.
The null state is an isolated fixed point.
The periods is same as those for rule-90,
though the detail behaviors are not (See Fig.~1 in paper I).}
\label{orbits4}
\end{figure}
\begin{figure}
\caption{The orbits of $N=5$ rule-90 cellular automaton.
This figure is the same as Fig.~2 in Paper I.}
\label{orbits5}
\end{figure}
\begin{table}
\caption{Periodicity of the transfer matrix $U$ for rule-90 and rule-150
cellular automata.}
\label{periodtable}
\begin{tabular}{r|ll}
N&rule-90&rule-150\\
\tableline
3&$U^3=0$&$U^4=I$\\
4&$U^6=I$&$U^6=I$\\
5&$U^5=U$&$U^5=U^4$\\
6&$U^{14}=I$&$U^{14}=I$\\
7&$U^7=7$&$U^8=I$\\
8&$U^{14}=I$&$U^{16}=U^2$\\
9&$U^{13}=U$&$U^{62}=I$\\
10&$U^{62}=I$&$U^{62}=I$\\
11&$U^{11}=U^3$&$U^{12}=U^8$\\
12&$U^{126}=I$&$U^{42}=I$\\
13&$U^{29}=U$&$U^{28}=I$\\
14&$U^{30}=I$&$U^{32}=U^2$\\
15&$U^{15}=7$&$U^{16}=I$\\
16&$U^{30}=I$&$U^{30}=I$\\
17&$U^{29}=U$&$U^{32}=U^4$\\
18&$U^{1022}=I$&$U^{1022}=I$\\
19&$U^{27}=U^3$&$U^{24}=I$\\
20&$U^{126}=I$&$U^{128}=U^2$\\
21&$U^{125}=U$&$U^{124}=I$\\
22&$U^{4094}=I$&$U^{4094}=I$\\
23&$U^{23}=U^7$&$U^{24}=U^{16}$\\
24&$U^{2046}=I$&$U^{2046}=I$\\
25&$U^{253}=U$&$U^{84}=I$\\
26&$U^{1022}=I$&$U^{1024}=U^2$\\
27&$U^{59}=U^3$&$U^{56}=I$\\
28&$U^{32766}=I$&$U^{32766}=I$\\
29&$U^{61}=U$&$U^{65}=U^5$\\
30&$U^{62}=I$&$U^{62}=I$\\
31&$U^{31}=0$&$U^{32}=I$\\
32&$U^{62}=I$&$U^{64}=U^2$\\
\end{tabular}
\end{table}
\narrowtext
\begin{table}
\caption{Eigenvalue polynomials for rule-90 cellular automata.}
\label{polynomial90}
\begin{tabular}{r|l}
$N$&$D^N(\lambda)$\\
\tableline
3&$\lambda^{3}$\\
4&$\lambda^{4}+\lambda^{2}+1$\\
5&$\lambda^{5}+\lambda$\\
6&$\lambda^{6}+\lambda^{4}+1$\\
7&$\lambda^{7}$\\
8&$\lambda^{8}+\lambda^{6}+\lambda^{4}+1$\\
9&$\lambda^{9}+\lambda^{5}+\lambda$\\
10&$\lambda^{10}+\lambda^{8}+\lambda^{4}+\lambda^{2}+1$\\
11&$\lambda^{11}+\lambda^{3}$\\
12&$\lambda^{12}+\lambda^{10}+\lambda^{8}+\lambda^{2}+1$\\
13&$\lambda^{13}+\lambda^{9}+\lambda$\\
14&$\lambda^{14}+\lambda^{12}+\lambda^{8}+1$\\
15&$\lambda^{15}$\\
16&$\lambda^{16}+\lambda^{14}+\lambda^{12}+\lambda^{8}+1$\\
17&$\lambda^{17}+\lambda^{13}+\lambda^{9}+\lambda$\\
18&$\lambda^{18}+\lambda^{16}+\lambda^{12}+\lambda^{10}+\lambda^{8}
+\lambda^{2}+1$\\
19&$\lambda^{19}+\lambda^{11}+\lambda^{3}$\\
20&$\lambda^{20}+\lambda^{18}+\lambda^{16}+\lambda^{10}+\lambda^{8}
+\lambda^{4}+\lambda^{2}+1$\\
21&$\lambda^{21}+\lambda^{17}+\lambda^{9}+\lambda^{5}+\lambda$\\
22&$\lambda^{22}+\lambda^{20}+\lambda^{16}+\lambda^{8}+\lambda^{6}
+\lambda^{4}+1$\\
23&$\lambda^{23}+\lambda^{7}$\\
24&$\lambda^{24}+\lambda^{22}+\lambda^{20}+\lambda^{16}+\lambda^{6}
+\lambda^{4}+1$\\
25&$\lambda^{25}+\lambda^{21}+\lambda^{17}+\lambda^{5}+\lambda$\\
26&$\lambda^{26}+\lambda^{24}+\lambda^{20}+\lambda^{18}+\lambda^{16}
+\lambda^{4}+\lambda^{2}+1$\\
27&$\lambda^{27}+\lambda^{19}+\lambda^{3}$\\
28&$\lambda^{28}+\lambda^{26}+\lambda^{24}+\lambda^{18}+\lambda^{16}
+\lambda^{2}+1$\\
29&$\lambda^{29}+\lambda^{25}+\lambda^{17}+\lambda$\\
30&$\lambda^{30}+\lambda^{28}+\lambda^{24}+\lambda^{16}+1$\\
31&$\lambda^{31}$\\
32&$\lambda^{32}+\lambda^{30}+\lambda^{28}+\lambda^{24}+\lambda^{16}+1$\\
\end{tabular}
\end{table}
\narrowtext
\begin{table}
\caption{Eigenvalue polynomials for rule-150 cellular automata.}
\label{polynomial150}
\begin{tabular}{r|l}
$N$&$D^N(\lambda)$\\
\tableline
3&$\lambda^{3}+\lambda^{2}+\lambda+1$\\
4&$\lambda^{4}+\lambda^{2}+1$\\
5&$\lambda^{5}+\lambda^{4}$\\
6&$\lambda^{6}+\lambda^{2}+1$\\
7&$\lambda^{7}+\lambda^{6}+\lambda^{5}+\lambda^{4}+\lambda^{3}
+\lambda^{2}+\lambda+1$\\
8&$\lambda^{8}+\lambda^{6}+\lambda^{2}$\\
9&$\lambda^{9}+\lambda^{8}+\lambda^{5}+\lambda^{4}+\lambda+1$\\
10&$\lambda^{10}+\lambda^{4}+1$\\
11&$\lambda^{11}+\lambda^{10}+\lambda^{9}+\lambda^{8}$\\
12&$\lambda^{12}+\lambda^{10}+\lambda^{8}+\lambda^{4}+1$\\
13&$\lambda^{13}+\lambda^{12}+\lambda^{5}+\lambda^{4}+\lambda+1$\\
14&$\lambda^{14}+\lambda^{10}+\lambda^{8}+\lambda^{6}+\lambda^{2}$\\
15&$\lambda^{15}+\lambda^{14}+\lambda^{13}+\lambda^{12}+\lambda^{11}
+\lambda^{10}+\lambda^{9}+\lambda^{8}+\lambda^{7}$\\
&\quad$+\lambda^{6}+\lambda^{5}+\lambda^{4}+\lambda^{3}
+\lambda^{2}+\lambda+1$\\
16&$\lambda^{16}+\lambda^{14}+\lambda^{10}+\lambda^{8}+\lambda^{6}
+\lambda^{2}+1$\\
17&$\lambda^{17}+\lambda^{16}+\lambda^{13}+\lambda^{12}+\lambda^{5}
+\lambda^{4}$\\
18&$\lambda^{18}+\lambda^{12}+\lambda^{10}+\lambda^{8}+\lambda^{4}
+\lambda^{2}+1$\\
19&$\lambda^{19}+\lambda^{18}+\lambda^{17}+\lambda^{16}+\lambda^{11}
+\lambda^{10}+\lambda^{9}$\\
&\quad$+\lambda^{8}+\lambda^{3}+\lambda^{2}+\lambda+1$\\
20&$\lambda^{20}+\lambda^{18}+\lambda^{16}+\lambda^{10}+\lambda^{2}$\\
21&$\lambda^{21}+\lambda^{20}+\lambda^{9}+\lambda^{8}+\lambda+1$\\
22&$\lambda^{22}+\lambda^{18}+\lambda^{16}+\lambda^{8}+1$\\
23&$\lambda^{23}+\lambda^{22}+\lambda^{21}+\lambda^{20}+\lambda^{19}
+\lambda^{18}+\lambda^{17}+\lambda^{16}$\\
24&$\lambda^{24}+\lambda^{22}+\lambda^{18}+\lambda^{8}+1$\\
25&$\lambda^{25}+\lambda^{24}+\lambda^{21}+\lambda^{20}+\lambda^{17}
+\lambda^{16}+\lambda^{9}+\lambda^{8}+\lambda+1$\\
26&$\lambda^{26}+\lambda^{20}+\lambda^{16}+\lambda^{10}+\lambda^{2}$\\
27&$\lambda^{27}+\lambda^{26}+\lambda^{25}+\lambda^{24}+\lambda^{11}
+\lambda^{10}+\lambda^{9}$\\
&\quad$+\lambda^{8}+\lambda^{3}+\lambda^{2}+\lambda+1$\\
28&$\lambda^{28}+\lambda^{26}+\lambda^{24}+\lambda^{20}+\lambda^{16}
+\lambda^{12}+\lambda^{10}+\lambda^{8}$\\
&\quad$+\lambda^{4}+\lambda^{2}+1$\\
29&$\lambda^{29}+\lambda^{28}+\lambda^{21}+\lambda^{20}+\lambda^{17}
+\lambda^{16}+\lambda^{13}+\lambda^{12}+\lambda^{5}+\lambda^{4}$\\
30&$\lambda^{30}+\lambda^{26}+\lambda^{24}+\lambda^{22}+\lambda^{18}
+\lambda^{14}+\lambda^{10}+\lambda^{8}$\\
&\quad$+\lambda^{6}+\lambda^{2}+1$\\
31&$\lambda^{31}+\lambda^{30}+\lambda^{29}+\lambda^{28}+\lambda^{27}
+\lambda^{26}+\lambda^{25}+\lambda^{24}+\lambda^{23}$\\
&\quad$+\lambda^{22}+\lambda^{21}+\lambda^{20}+\lambda^{19}
+\lambda^{18}+\lambda^{17}+\lambda^{16}+\lambda^{15}$\\
&\quad\quad$+\lambda^{14}+\lambda^{13}+\lambda^{12}+\lambda^{11}
+\lambda^{10}+\lambda^{9}+\lambda^{8}+\lambda^{7}$\\
&\quad\quad\quad$+\lambda^{6}+\lambda^{5}+\lambda^{4}
+\lambda^{3}+\lambda^{2}+\lambda+1$\\
32&$\lambda^{32}+\lambda^{30}+\lambda^{26}+\lambda^{24}+\lambda^{22}
+\lambda^{18}+\lambda^{14}+\lambda^{10}$\\
&\quad$+\lambda^{8}+\lambda^{6}+\lambda^{2}$\\
\end{tabular}
\end{table}
%
\end{document}